\documentclass[sigplan]{acmart}
\renewcommand\footnotetextcopyrightpermission[1]{} % removes footnote with conference information in first column
\makeatletter
\renewcommand\@formatdoi[1]{\ignorespaces}
\makeatother

\usepackage{fancyhdr}
\usepackage{booktabs} % For formal tables
\usepackage{xcolor}
\definecolor{pmGreen}{RGB}{0,156,121}
\definecolor{eqblue}{RGB}{74, 5, 119}
\definecolor{ocre}{RGB}{128,102,25}
\everymath{\color{eqblue}}
\usepackage[format=hang,labelfont={bf, color=black}, font={small, color=ocre}]{caption}
\usepackage{graphicx, subfig}
\usepackage{amsmath}
\usepackage[ruled,vlined,linesnumbered]{algorithm2e}
\usepackage{enumitem}

\usepackage{listings}
\definecolor{dkgray}{rgb}{0.5,0.5,0.5}
\definecolor{gray}{rgb}{0.5,0.5,0.5}
\definecolor{mauve}{rgb}{0.58,0,0.82}
\setlist[itemize]{leftmargin=*}

\begin{document}
\title{Automatic acceleration of Numpy applications on GPUs and
  multicore CPUs}
%% \subtitle{Extended Abstract}
%% \subtitlenote{The full version of the author's guide is available as
  %% \texttt{acmart.pdf} document}

\author{Mahesh Ravishankar}
%% \authornote{Dr.~Trovato insisted his name be first.}
%% \orcid{1234-5678-9012}
\affiliation{%
  \institution{NVIDIA Corporation}
  \streetaddress{11431 NE Willows Road}
  \city{Redmond}
  \state{Washington}
  \postcode{98052}
}
\email{mravishankar@nvidia.com}

\author{Vinod Grover}
\affiliation{%
  \institution{NVIDIA Corporation}
  \streetaddress{11431 NE Willows Road}
  \city{Redmond}
  \state{Washington}
  \postcode{98052}
}
\email{vgrover@nvidia.com}

%% % The default list of authors is too long for headers.
%% \renewcommand{\shortauthors}{M. Ravishankar et. al.}

\begin{abstract}
Frameworks like Numpy are a popular choice for application developers
from varied fields such as image processing to bio-informatics to
machine learning. Numpy is often used to develop prototypes or for
deployment since it provides efficient implementation for operations
involving arrays. Such an approach requires every operation to be
executed eagerly. The result of each operation needs to be stored in
memory which increases the memory footprint of the application. It
also increases the bandwidth requirements since all uses must read
from this memory. We propose an approach that records the sequence of
Numpy operations for defered execution. When the values of an array
are needed, for example when the values are stored to disk or
displayed on screen, the sequence of operations required to compute
these value are compiled into a function and executed. This removes
the need to store/load intermediates in slow memory, resulting in
better performance. In cases where the library implementation is more
efficient (like matrix-matrix multiply), those are used instead. The
approach also allows us to seamlessly target both multicore CPUs and
NVIDIA GPUs, thereby porting the Numpy application to these
architectures without changing the user program. The benefit of the
approach is evaluated by targeting computation samples from various
domains and on average on order of magnitude performance improvement
over Numpy is observed.

\end{abstract}

%% \begin{CCSXML}
%% <ccs2012>
%%  <concept>
%%   <concept_id>10010520.10010553.10010562</concept_id>
%%   <concept_desc>Computer systems organization~Embedded systems</concept_desc>
%%   <concept_significance>500</concept_significance>
%%  </concept>
%%  <concept>
%%   <concept_id>10010520.10010575.10010755</concept_id>
%%   <concept_desc>Computer systems organization~Redundancy</concept_desc>
%%   <concept_significance>300</concept_significance>
%%  </concept>
%%  <concept>
%%   <concept_id>10010520.10010553.10010554</concept_id>
%%   <concept_desc>Computer systems organization~Robotics</concept_desc>
%%   <concept_significance>100</concept_significance>
%%  </concept>
%%  <concept>
%%   <concept_id>10003033.10003083.10003095</concept_id>
%%   <concept_desc>Networks~Network reliability</concept_desc>
%%   <concept_significance>100</concept_significance>
%%  </concept>
%% </ccs2012>
%% \end{CCSXML}

%% \ccsdesc[500]{Computer systems organization~Embedded systems}
%% \ccsdesc[300]{Computer systems organization~Redundancy}
%% \ccsdesc{Computer systems organization~Robotics}
%% \ccsdesc[100]{Networks~Network reliability}

\keywords{Numpy, GPUs, Machine Learning}

\maketitle

\section{Introduction}
\label{sec:introduction}

Numpy~\cite{numpy} has long been the default module for developing
applications in Python that rely on array-based data structures. Image
processing, deep learning, bio-informatics are just a few domains
where Python and Numpy are used to either develop prototypes or
applications that are deployed for public use. The main reason for
this is that Numpy provides efficient implementation for operations
involving arrays. At the same time, Numpy has several
limitations. Firstly, efficient implementations are provided for CPU
execution, i.e., there is no off the shelf method to execute the Numpy
operation on the GPU. A majority of the operations provided by Numpy,
like \texttt{numpy.add} or \texttt{numpy.subtract} are embarrassingly
parallel, i.e., The same computation is executed independently on each
element of the array. Such a computation pattern matches well with the
GPU execution model which requires the use of lots of parallel threads
with minimal communication for efficient execution. For more complex
operations like \texttt{np.dot}, which implements matrix-matrix and
matrix-vector multiplications, existing libraries provide efficient
implementation on GPUs. For example, cuBLAS~\cite{cublas} provides
efficient GEMM implementation for NVIDIA GPUs. Yet, there is no
mechanism in Numpy to leverage these libraries.

\begin{figure}
\begin{lstlisting}[caption={Numpy implementation of simple arithmetic
    ops}, label={lst:example}]
import numpy as np
x = np.array([10, 20], np.int32)
y = np.array([20], np.int32)
z = x*x + 2*x*y + y*y
print(z)
\end{lstlisting}
\vspace{-2em}
\end{figure}

A second limitation comes from the fact that Numpy is a library. Each
operation provided by Numpy is implemented through a dedicated
function. For example, consider the snippet shown in
Listing~\ref{lst:example}. The four multiplications and two additions
are done eagerly with intermediate result (which in itself is an
array), stored in memory. These are read again during its subsequent
uses. Since the actual computation is simple, the entire execution is
bandwidth bound.  As the arrays get larger, streaming the array from
memory multiple times results in poor performance. The overhead of
memory allocation and deallocation of temporary arrays further worsens
the situation.

In this paper, we propose a framework called Grumpy, that employs an
alternative approach that addresses both of these issues. Instead of
performing each operation eagerly, we build a representation of the
operations seen so far. This is done by providing a python module
called \texttt{grumpy} that can be used instead of numpy. It provides
the same operations as \texttt{numpy} and can be used as a drop-in
replacement for it. When the result of the operation is needed (due to
the \texttt{print} operation, for example), we use a JIT compiler to
generate code for all the operations seen so far and execute it as a
single function.

This approach has the following advantages
\begin{itemize}
\item There is no need to explicitly store the intermediate results in
  memory. For example, for Listing~\ref{lst:example}, the generated
  function directly computes the value of \texttt{z} without storing
  the values of all the intermediate arrays in memory.  
\item Since the JIT compiler can see more of the computation than
  individual operations, traditional compiler optimizations can now be
  applied to generate a more efficient function.
\item The JIT compiler can generate native code to target multiple
  architectures. By using the embarrassingly parallel nature of the
  array operations in numpy, we can automatically generate code and
  run the computation either on GPUs (specifically NVIDIA GPUs) or on
  multi-core CPUs.
\end{itemize}

In general, it is not always efficient to do all evaluation lazily, or
to generated code for all operations using a JIT compiler. For
example, consider the snippet in Listing~\ref{lst:mnist} which
representative of MNIST computation from Deep Learning domain. It
contains operations such as transpose and matrix-vector
multiplies. Instead of generating code using a JIT compiler, it is
more efficient to use tuned implementations provided by libraries such
as LAPACK~\cite{lapack} or cuBLAS~\cite{cublas}. To account for this
Grumpy recognizes computations that are better offloaded to efficient
library implementation, while generating optimized fused kernels for
the rest of the computation.

\begin{figure}
\begin{lstlisting}[caption={Simple MNIST in Numpy},
  label={lst:mnist}]
import numpy as np
W = np.array([10, 200], np.float32);
a = np.array([200], np.float32);
b = np.array([10], np.float32);
x = np.transpose(W).dot(max(a, 0)) + b;
output = 1.0 / (1.0 + np.exp(-x));
print(output);
\end{lstlisting}
\vspace{-2em}
\end{figure}

The rest of the paper is organized as follows. Section~\ref{sec:fusion}
describes the internal representation used to describe the
computation, and how this is used to offload parts of the computation
to a library while the rest of the computation is JIT
compiled. Section~\ref{sec:codegen} describes the code-generation
process within the JIT compiler. Section~\ref{sec:results}
evaluates the performance improvements from using this approach on a
few benchmarks. Section~\ref{sec:related} describes other frameworks
that have similar goals and describes the advantages/disadvantages of
the Grumpy compared with these frameworks.

\section{Fusion of operations in the DAG}
\label{sec:fusion}

\begin{figure}
  \centering
  \includegraphics[scale=0.4]{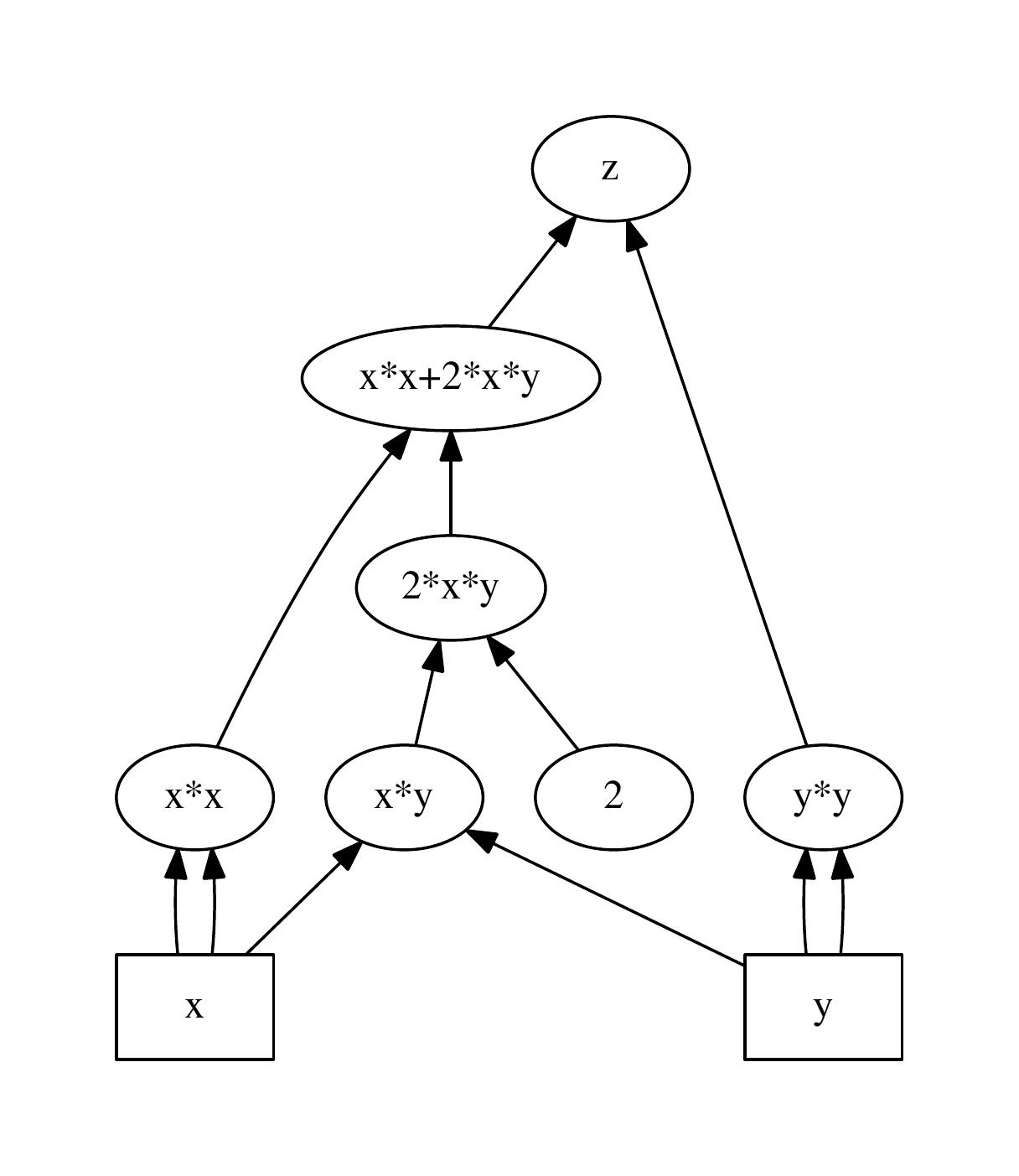}
  \caption{DAG representation of the computation in Listing~\ref{lst:example}}
  \label{fig:example}
  \vspace{-1em}
\end{figure}

To move from an eager evaluation approach to a lazy evaluation
approach we need a representation of the computation seen so far. A
Directed Acyclic Graph, $G = (V, E)$ is a natural abstraction to
represent the sequence of operations to be evaluated, and the data
dependencies between them. A node $n \in V$ represent the result of an
operation, which is in general an $n$-dimensional tensor. An edge $e
\in E$ represents the use of the tensor produced by the source node in
the target node. The leaves of the graph represent inputs to the
computation. The root is the result of the computation. It is not
required to store the intermediates in memory, but it is required to
produce the output in memory. Similar to numpy, the size of the tensor
represented by each node, and the data type of the tensor elements are
computed automatically while the DAG is built based on the semantics
of each operation. Figure~\ref{fig:example} shows the DAG that
represents the computation in
Listing~\ref{lst:example}. \texttt{output} is the result of the
computation and is therefore the root. \texttt{W}, \texttt{a} and
\texttt{b} are inputs to the computation and are the leaves of the
DAG.

The nodes in the graph can be used to represent any mathematical
operation ranging from addition, subtraction of arrays to more complex
operations like,

\begin{itemize}
\item BLAS computations such as matrix-vector multiply or
  matrix-matrix multiply
\item Reduction and scan along multiple (or all) dimension of the
  array.
\item Reads and writes of portion of array through slicing operations.
\item Index and layout transformation operations like reshape and
  multi-dimensional transpose operations.
\end{itemize}

to name a few. The semantics of all operations match the semantics of
the corresponding numpy operations, including aspects like
broadcast~\cite{broadcast} for pointwise operations. Note that we do not
need to account for branching constructs like \emph{if-then-else} or
loops. Since Python is an interpreted language, only the operations
executed are seen by the Grumpy module, and needs to be represented in
the DAG.

Each node in the graph has three attributes
\begin{itemize}
\item \emph{operation}: Represents the operations being represented by
  the node
\item \emph{data}: Contains the values in the $n$-dimensional tensor
  that would be generated after the operation represented by this node
  is executed.
\item \emph{is\_materialized} : A boolean attribute that signifies if
  the \emph{data} has been computed and stored in memory or not.
\end{itemize}

At the start of the computation, only the inputs nodes have the
\emph{data} field non-empty and represents the values in the arrays
that are inputs to the computation. The \emph{is\_materialized} field
is also \texttt{true} only for the inputs. As more operations are
encountered, the DAG is built to capture the semantics of the
operations specified for lazy evaluation. The \emph{is\_materialized}
field is set to \texttt{false} for these nodes and the \emph{data}
field is empty. If the result of a particular node is needed (through
a print or other similar operations), the DAG is traversed backwards
to find all the operations needed to evaluate the result. On
encountering a materialized node (one with the \emph{is\_materialized}
field set to \texttt{true}), none of its predecessors are visited as
part of the traversal.  At the end of the traversal, all visited nodes
form a subgraph, which itself is a DAG where the leaves, and only the
leaves, are materialized. This subDAG is passed to the JIT compiler
that compiles the graph into a single fused function, which accepts
the input arrays as arguments, and the generates the array containing
the result of the operation at the
root.. Listing~\ref{lst:example-code} is representative of the code
generated for the subDAG rooted at \texttt{z} with leaves \texttt{x}
and \texttt{y} in Figure~\ref{fig:example}. Note that the generated
code correctly implements the broadcast semantics of multiplying
arrays \texttt{x} and \texttt{y}. Details of code-generation are
presented in Section~\ref{sec:codegen}.

\begin{figure}
\begin{lstlisting}[language=c, caption={Representative function that
      fuses the computation in Listing~\ref{lst:example}},
    label={lst:example-code}, rangeprefix={\/\/\ \{\ },
    rangesuffix={\ \}},
    linerange=targetcodegenstart-targetcodegenend]
// { targetcodegenstart }
void fused_example(const int &x[10][20], const int &y[20], int &z[10][20]) {
  for (int i = 0; i < 10; i++)
    for (int j = 0; j < 20; j++)
      z[i][j] = x[i][j] * x[i][j] + 2 * x[i][j] * y[j] + y[j] * y[j];
}
// { targetcodegenend }

// { lambdastart }
void f(int i, int j, const int *x, const int *y, int *z) {
  z[i*20+j] = x[i*20+j]*x[i*20+j] + 2*x[i*20+j]*y[j] + y[j]*y[j];
}
// { lambdaend }
\end{lstlisting}
\vspace{-2em}
\end{figure}

The JIT compiler in Grumpy compiles this code to execute on a
multi-core CPU or NVIDIA GPU. On execution, the result of the output
is stored in memory. The attribute \emph{data} of this node is
associated with these result array stored in memory, and the node is
set as been materialized. Since Python is an interpreted language, at
this point it is not known how the values computed for this node will
be used. Marking the node as materialized ensures that other nodes
which depend on this node will use the value computed in memory,
instead of recomputing these values.

As mentioned in Section~\ref{sec:introduction}, for some computations
(like the matrix-vector multiply in Listing~\ref{lst:mnist}) it is
better to fall back to more efficient library implementation than to
have the JIT compiler generate code for such operations. To enable
this, when such an operation is encountered, all the operands to this
operation are materialized. This is done since all the operands would
have been computed and stored in memory, for use within the library
function invoked to compute the result of the operation. On return
from the library call, the node corresponding to the operation is
marked as materialized so that all uses of this node will use the
values generated from the library call.

\begin{figure}
  \centering
  \includegraphics[scale=0.4]{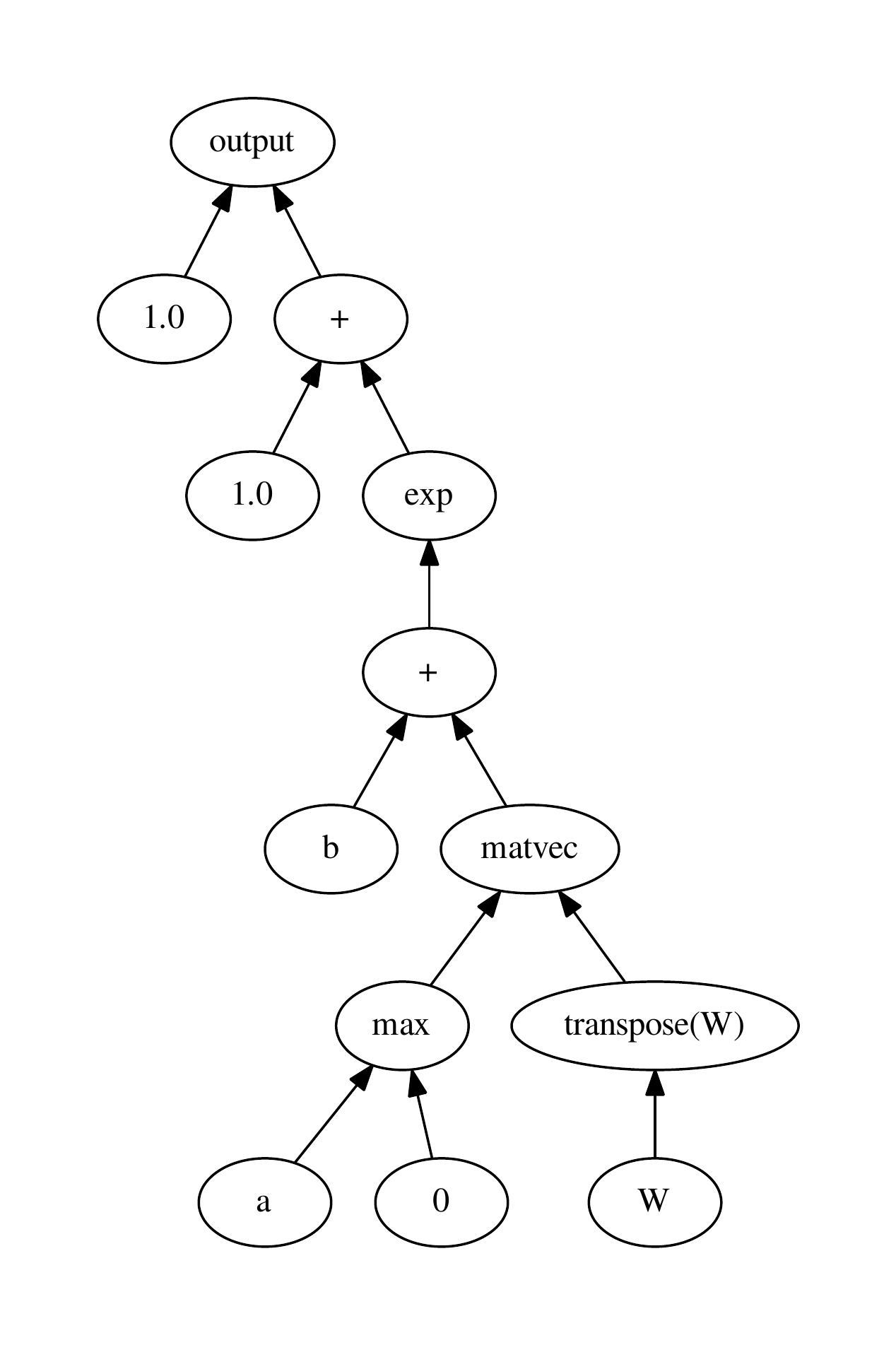}
  \caption{DAG representation of MNIST computation in Listing~\ref{lst:mnist}}
  \label{fig:mnist}
  \vspace{-1em}
\end{figure}

To illustrate the above mechanism, consider the DAG for the MNIST
example from Listing~\ref{lst:mnist} shown in
Figure~\ref{fig:mnist}. The operation for node \texttt{matvec} is
better handled by cuBLAS for GPU execution, by invoking the relevant
\emph{gemv} API calls. Before invoking the function, the values of the
matrix and the vector (which are the operands to matvec) need to be
materialized. The value of \texttt{max} is computed by generating a
function to implement the max function, JIT-compiling it and executing
it. cuBLAS also provides a version of \emph{gemv} computation where
the input matrix is stored as transpose. This is recognized by Grumpy,
and it proceeds to materialize the operand of the transpose. Since
this is the array \texttt{W} that is already materialized, there is
nothing to do here. On return from the cuBLAS function, the
\texttt{matvec} node is marked as materialized. The rest of the
operations to compute the \texttt{output} can be fused into a single
function with \texttt{matvec} and \texttt{b} being inputs to the
function.

\begin{algorithm}
  \footnotesize
  \SetInd{0.5em}{0.5em}
  \SetKwInOut{Input}{Input}
  \SetKwInOut{InOut}{InOut}
  \Input {
    \textbf{root} : Root node of the subGraph to be created
  }
  \InOut {
    \textbf{graphList} : List of subGraphs, each either executed by a
    library or is JIT-Compiled. \\
    \textbf{visited} : Set of nodes visited
  }
  \Begin
    {
      newGraph = $\phi$ \;
      candidates = $\phi$ \;
      \If {root $\notin$ visited} {
        candidates.push(root) \;
        visited = visited $\cup$ root \;
        set\_materialized(node) \;
      }
      \While{$\lnot$ is\_empty(candidates)} {
        node = candidates.pop() \;
        newGraph.nodes = newGraph.nodes $\cup$ node \;
        \If {$\lnot$ is\_materialized(node)} {
          \If {materialize\_node(node) } {
            create\_subgraph(node, graphList, visited) \;
          }
          \ForEach {pred $\in$ predecessors(node)} {
            \If {pred $\notin$ visited} {
              \If {materialize\_pred\_of\_node(node, pred)} {
                create\_subgraph(pred, graphList, visited) \;
                newGraph.nodes = newGraph.nodes $\cup$ pred \;
              } \ElseIf {$\lnot$ is\_materialized(pred)} {
                visited = visited $\cup$ pred \;
                candidates.push(pred) \;
              }
            } \ElseIf {pred $\notin$ newGraph.nodes} {
              newVisited = $\phi$ \label{algoline:multipleUse} \;
              create\_subgraph(pred, graphList, newVisited) \;
            }
            newGraph.edges = newGraph.edges $\cup$ \{node, pred\} \;
          }
        }
      }
      graphList.append(newGraph) \;
    }
    \caption{create\_subgraph(rootNodes, graphList, visited)}
    \label{algo:create_subgraph}
\end{algorithm}

The algorithm used to split an input DAG into a list of sub-graphs,
each of which is to be either JIT compiled and executed or handed off
to a library is shown in Algorithm~\ref{algo:create_subgraph}. The
sub-graphs are to be executed in the order specified by this list. It
uses two helper functions to create the sub-graphs,
\texttt{materialize\_node} and \texttt{materialize\_pred\_of\_node}.

\texttt{materialize\_node} is a function that checks if a node has to
be materialized based on the properties of the operation. Some of the
cases where this function returns \texttt{true} are listed below.
\begin{itemize}
\item When the operation is better handled through use of libraries,
  like cuBLAS.
\item When the operation is not embarrassingly parallel, like
  reductions or scans (along one or many dimensions of the
  array). These operations require inter-thread communication. When
  executed in parallel, fusing these operations with their successors
  one would either need a explicit synchronization between threads or
  these computations have to be executed redundantly in all the
  threads. Not all hardware have effective mechanism to do the former
  (for example on NVIDIA GPUs it is not always possible to synchronize
  threads across all thread blocks) and the latter is not efficient.
\end{itemize}

\texttt{materialize\_pred\_of\_node} is a function that returns
\texttt{true} if a predecessors of a node has to be materialized. This
function returns \texttt{true} for cases where the node represents an
operation that is to be executed by a library. Since the operands of
the library have to be computed and stored in memory before the
library call, this step is required. Note that sometimes it is better
to not materialize the immediate predecessor. For example, for an
matrix-matrix multiply or a matrix-vector multiply, if the predecessor
is a transpose, cuBLAS provides methods that allow the input to be
transposed before the computation is performed. In this case it is
better to materialize the predecessor of the transpose. This detail is
omitted from the Algorithm~\ref{algo:create_subgraph} for sake of
clarity.

Another aspect of algorithm to note is in
lines~\ref{algoline:multipleUse}. If a node encountered has been
visited during the subgraph creation, but does not belong to the
current subgraph, then it is a node that belongs to a subgraph that
was created earlier. One way to handle this is to include all the
nodes in the sub-DAG rooted at this node into the current
subgraph. This would mean the computation represent by this sub-DAG is
replicated in all subgraphs that use the result of this node. An
alternative approach is to materialize this node so that it becomes a
leaf to all subgraphs that use this node. This is the approach used in
Grumpy.

To control JIT compilation overhead, when the size of the subDAG
rooted as each node reaches a threshold it is JIT compiled. The
heuristics used here are outside the scope of this paper.

\section{Code Generation for fused operations}
\label{sec:codegen}

The JIT compiler used within Grumpy is based on
LambdaJIT~\cite{lambdajit}. It is an LLVM based compiler that can
generate code to execute either on the multicore CPU or on the
GPU. Its supports code generation for three primitive operations:
\emph{map}, \emph{map-reduce} and \emph{map-scan}. This section
provides details of the code-generation process.

\subsection{Map operations}
\label{subsec:map-codegen}

Map operations are a natural representation of computations where the
same operations is to be performed on a different elements of an
array. These operations can be done in parallel without any need for
synchronization. For example, consider the code in
Listing~\ref{lst:example-code} that implements the computation
represented by the DAG in Figure~\ref{fig:example}. This computation
can be viewed as executing the function \texttt{example\_pt\_fn} shown
in Listing~\ref{lst:example-lambda} over a 2D orthogonal iteration
space. Given this function (refered to as point function) and the
iteration space dimensions, the JIT compiler will generate code that
partitions this iteration space to either
\begin{itemize}
\item CPU threads on multicore CPUs with each thread executing a block
  of this iteration space, or
\item GPU with each thread executing one point in this iteration space
\end{itemize}

\begin{figure}
\begin{lstlisting}[language=c, caption={Map function that is used to
      generate the fused code similar to
      Listing~\ref{lst:example-code}}, label={lst:example-lambda},
    rangeprefix={\/\/\ \{\ }, rangesuffix={\ \}},
    linerange=lambdastart-lambdaend]
// { targetcodegenstart }
void fused_example(const int &x[10][20], const int &y[20], int &z[10][20]) {
  for (int i = 0; i < 10; i++)
    for (int j = 0; j < 20; j++)
      z[i][j] = x[i][j] * x[i][j] + 2 * x[i][j] * y[j] + y[j] * y[j];
}
// { targetcodegenend }

// { lambdastart }
void f(int i, int j, const int *x, const int *y, int *z) {
  z[i*20+j] = x[i*20+j]*x[i*20+j] + 2*x[i*20+j]*y[j] + y[j]*y[j];
}
// { lambdaend }
\end{lstlisting}
  \vspace{-2em}
\end{figure}

In general the code-generator accepts a point function and an
$n$-Dimensional iteration space to generate the code for the parallel
execution of the map operation. It is expected that the first $n$
arguments of point function are integers that represent the point in
the iteration space being executed.

To generate the point function for a subgraph found in
Algorithm~\ref{algo:create_subgraph}, the code generation starts from
the root of the subgraph. The shape of the result tensor (i.e., the
result of the operation represented by the root node) gives the
dimensionality and shape of the iteration space to be used in the
generated code. The first $n$ arguments of the point function also
represent the index into the result tensor where the value computed by
the point function is to be stored. These indices are translated into
indices for predecessors, and so on till the leaves of the subgraph
are reached. The index expression at the leaves are used to reads
inputs to the subgraph (Algorithm~\ref{algo:create_subgraph} ensures
that the leaves are materialized). For example, consider a subgraph
with a transpose operation of a 2D tensor as the root. Say, the first
two arguments of the point function are $i$ and $j$, i.e., the point
function computes the value of element $(i,j)$. Then the indices to
use for the predecessor would be $(j, i)$. The index expression
translation needs to account for broadcast semantics, slicing, and
other index transformation operations. The point function generated
and the shape of the root node are passed to a routine that generates
code for map operation. This function, when invoked, executes the
computation represented by the subgraph.

\subsection{Map-Reduce and Map-Scan operations}
\label{subsec:mapreduce-codegen}

The \emph{map-reduce} operations represents the computation where
elements of a tensor are combined to produce a single value using an
associative and commutative operation (we assume floating point
operations satisfy both within some error bounds). For example, the
\texttt{numpy.sum()} operation, which is a reduction with sum operator
over all axis of the input array fits such an abstraction. To generate
code for such computations, the code generators takes two functions as
inputs. First, a binary operation that takes two values of the same
type and combines them using an associative and commutative operation
to generate a result of the same type. The second is a map function,
similar to the function describes in Section~\ref{subsec:map-codegen},
but instead of \texttt{void} return type, it returns a value that is
of the same type as the arguments of the binary operator. The map
function is used to produce all the elements that are to be combined
using the binary operator to produce a single final result value. For
example, the Listing~\ref{lst:innerproduct} shows the map function,
\texttt{ip\_map} and the binary operator, \texttt{ip\_reduce} to be
used to implement an inner-product of two $1$-D vectors \texttt{a} and
\texttt{b}. The map operations first performs the element wise
multiplication of the two vectors to generate the values that are to
be combined using \texttt{+} operation to get the final result.

\begin{figure}
\begin{lstlisting}[language=c, caption={Map and Reduce function used
      to generate the map-reduce function that computes the inner
      product of two $1$-D vectors}, label={lst:innerproduct},
    rangeprefix={\/\/\ \{\ }, rangesuffix={\ \}},
    linerange=start-end]
// { start }
float ip_map(int i, float *a, float *b) {
  return a[i] * b[i];
}
float ip_reduce(float a, float b) {
  return a + b;
}
// { end }
\end{lstlisting}
\vspace{-2em}
\end{figure}

When the subgraph to be JIT compiled has a root which represents a
reduction value to generate a single value, it first generates the
binary function based on the reduction operation being used. It then
generates the point function for the map in a process similar to the
one described in Section~\ref{subsec:map-codegen}. The dimensionality
of the map function is same as that of the operand of the root node,
since its the result of the operand that are the values to be combined
during the reduction operation represented by the root. The shape of
the operand to the root node, the map function and the binary
operation is used to the map-reduce code for the desired target.

On CPU the iteration space for the map operation is split amongst the
threads in a blocked fashion. The result of the map operation are then
combined using the binary operator to get a value per thread. The
values generated for each thread is then combined using the binary
operator to get the final result.

On the GPU, the iteration space of the map operation is distributed
amongst threads, with one point per thread. The result generated from
threads in a thread block are combined in shared memory to a single
value. The result from each thread block is written out to global
memory. A separate kernel is used to do the final reduction of these
values to get the final result. Note that for this kernel does not use
the map operation. It only uses the binary operation to combine these
values.

For cases where the root of the subgraph that is to be JIT compiled is
a scan operation, a map-scan primitive is used. It requires the same
map point function and binary operator as above. Further details about
the mechanism to compute the map-scan in parallel on the CPU and GPU
are not relevant to this paper. For further details about the scan
algorithm generated please refer to the work-efficient parallel scan
algorithm described here~\cite{gpuscan}.

\section{Experimental Results}
\label{sec:results}

The Grumpy framework described in Sections~\ref{sec:fusion}
and~\ref{sec:codegen} is deployed as a python module, called
\texttt{grumpy} that implements the same interface as Numpy. The goal
is that the only change a user has to make is to import this module
instead of \texttt{numpy}. While we would like handle all the numpy
methods within Grumpy, it is a daunting task due to sheer number of
methods that Numpy supports. To ensure we achieve the desired
deployment model we ensure that \texttt{grumpy} and \texttt{numpy}
modules interact seamlessly. Internally, the \texttt{grumpy} module
intercepts operation sequence that can handled by the framework and
passes it on to the core Grumpy libraries. For the unhandled
operations, it is forwarded to the \texttt{numpy} module. We describe
briefly few aspects of this mechanism.

\subsection{\texttt{grumpy} - \texttt{numpy} Interoperability}
\label{subsec:grumpy_numpy}

The \texttt{grumpy} python module provides an \texttt{ndarray} object
similar to the one that is provided by \texttt{numpy}. This object has
the same methods, attributes, etc. as Numpy version of the same
object. While it is not implemented this way, it can be thought of as
a derived object of the \texttt{numpy.ndarray}. An instance of the
\texttt{grumpy.ndarray} object represents a node in the DAG internal
representation in Grumpy. To be able to inter-operate with
\texttt{numpy} seamlessly, we have to ensure that each method provided
by \texttt{grumpy} knows how to handle input arguments being
\texttt{numpy} arrays. These typically form the leaves of the sequence
of operations that are then evaluated lazily using Grumpy, and
therefore act as inputs to the computation. If an unsupported
operation is encountered (which would include print methods), before
falling back to \texttt{numpy}, the nodes that correspond to the
arrays being passed to the fallback routine need to be
materialized. This triggers the materialization process, and the
result is forwarded to the numpy methods as a numpy array.

\subsection{Performance gains on multi-core CPUs and GPU}
\label{subsec:perf}

\begin{figure}
\begin{lstlisting}[caption={Jacobi stencil in Numpy},
    label={lst:jacobi}]
center = grid[1:-1, 1:-1]
north  = grid[0:-2, 1:-1]
east   = grid[1:-1, 2:  ]
west   = grid[1:-1, 0:-2]
south  = grid[2:  , 1:-1]
for i in range(iterations):
    center[:] = 0.2*(center+north+east+west+south)
return grid
\end{lstlisting}
\vspace{-2em}
\end{figure}

\begin{figure*}
  \centering \subfloat[][BlackScholes]{
    \includegraphics[scale=0.7]{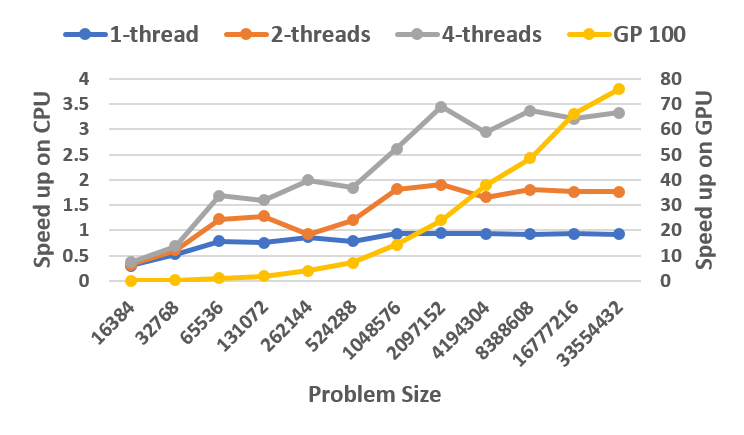}
    \label{fig:blackscholes}
  }
  \subfloat[][KMeans]{
    \includegraphics[scale=0.7]{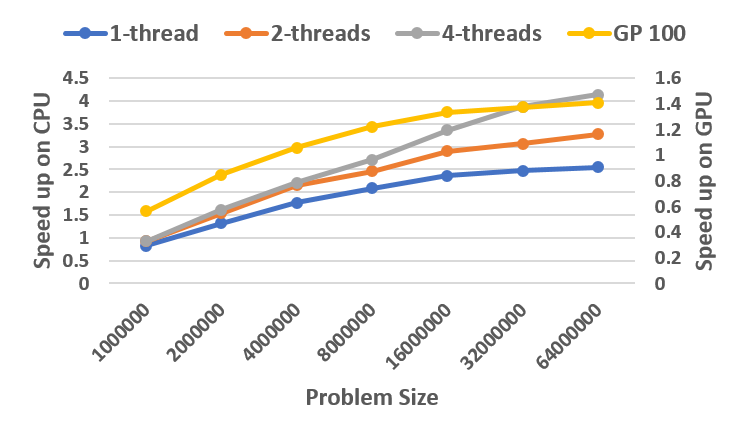}
    \label{fig:kmeans}
  }
  \\
  \subfloat[][Jacobian Stencil]{
    \includegraphics[scale=0.7]{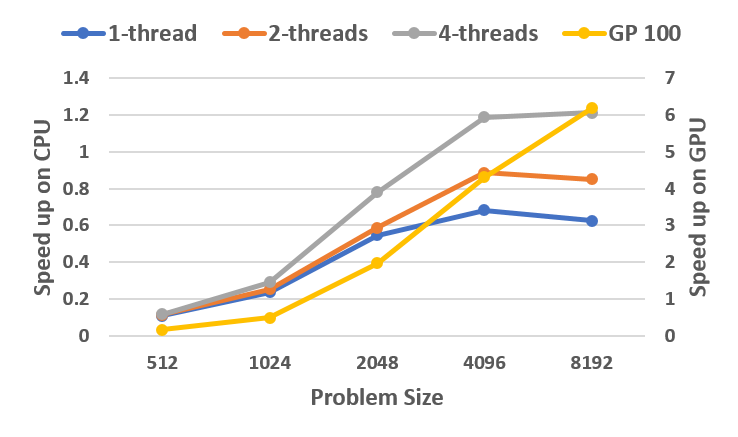}
    \label{fig:jacobi}
  }
  %% \subfloat[][Closest Distance on Sphere]{
  %%   \includegraphics[scale=0.7]{figs/cds.png}
  %%   \label{fig:cds}
  %% }
  %% \\
  \subfloat[][RNN]{
    \includegraphics[scale=0.7]{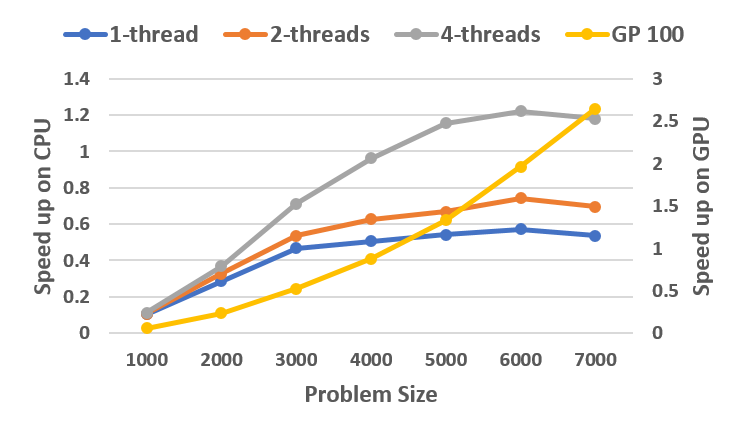}
    \label{fig:rnn}
  }
  \caption{Performance comparison against Numpy on multicore CPU and NVIDIA GPU}
  \label{fig:perfbase}
\end{figure*}

For the rest of the section we discuss some of the benchmarks that
were targeted using this approach. We picked six benchmarks from
different fields for evaluation. Black-Scholes is a widely used
computation in financial application domain. KMeans is a basic
clustering algorithm from statistical analysis domain. Jacobi stencil
represents a stencil computation used in many simulation
applications. Listing~\ref{lst:jacobi} so the numpy implementation
used for the baseline. The fourth application is from deep-learning
domain and is an RNN training sample. The Grumpy JIT-Compiler was
implemented in LLVM-3.8. On the CPU, we can use LLVM to JIT-Compile to
\texttt{x86} code. On the GPU LLVM produces PTX~\cite{ptx}
instructions, that are similar to assembly instructions. The
CUDA~\cite{cuda} driver can compile the PTX to the machine
instructions. We use CUDA-9.0 for our experiments. To support GPU
execution, inputs to the kernel to be executed are automatically
copied onto the device and the results are copied back to the
host. Our implementation does not implement any specific CPU or GPU
optimization to reduce the complexity of handling multiple
backends. For example, we do not use vectorization for CPUs. We do not
make use of fast shared memory on GPUs to improve performance. As a
result the main benefit from using Grumpy would be only through fusion
or through use of specialized libraries for certain operations.

Figure~\ref{fig:perfbase} shows the performance improvements over Numpy
execution when using Grumpy. For each individual figure the speedup
over Numpy when using $1$, $2$ and $4$ CPU threads are shown (these
lines use values from the left side y-axis). The figure also shows the
speed up when executing the computation using a GP 100 (using the
value from the right side y-axis).The performance numbers include the
overhead of JIT compilation. It is expected that for small problem
sizes, the overhead of JIT compilation dominates. As a result it
should be expected that Grumpy would be considerably slower than
Numpy. As problem size increases and the computation time dominates the
overall execution time, the performance benefits of fusion in Grumpy
as well as the ability to execute the computation in parallel on CPUs
or to offload to GPUs starts showing dividends. We present below on
analysis of the performance improvements on each of these benchmarks.

Black-Scholes (Figure~\ref{fig:blackscholes}) is a computation that
contain a lot of point-wise operations which could be completely
fused. For small problem sizes, the JIT compilation overhead
dominates, but for large problem sizes, Grumpy provides as much as
$76$x speed up over Numpy. No hand tuned kernels from libraries are
needed for executing this computation. KMeans computation is
inherently poorly suited for the GPU execution model due to irregular
accesses to the input data during clustering. For this computation as
well, there are no computations offloaded to libraries. Due to these
factors, the GPU performance (Figure~\ref{fig:kmeans}) is not as
drastic from Grumpy. Interestingly, for large problem sizes, the
multi-core CPU performance shows good improvement, both due to fusion
and parallelism.

The Jacobian stencil computation shows that Grumpy is able to fuse the
computation into a single kernel even when slicing. The generated code
is what one would have written in a low-language like CUDA. An
interesting point to note is that slicing in Numpy doesn't create a new
array, but a view of the same buffer. That is the main reason behind
almost no improvement on CPU from fusion. Further, being bandwidth
bound, parallelism doesn't help as much on the CPU. GPU performance
still shows about $6$ times improvement over Numpy.

The RNN training sample has a lot of dot products, and very few
sequences of point functions. This results in the compilation overhead
having a major impact on the performance. Consequently, Grumpy obtains
a speed up of just $2.5$ times over Numpy when running on GPUs. On
CPUs no improvement is seen. One issue here is specific to the current
implementation. Grumpy doesn't use BLAS libraries for CPUs. Instead it
uses a rudimentary matrix-matrix multiply kernel which is not
optimized for cache or to use vector instruction units on CPUs. Since
matrix-matrix multiplies form a major portion of the computation, this
implementation deficiency hurts the speed up achieved with Grumpy on
CPUs

\begin{figure*}
  \centering \subfloat[][BlackScholes]{
    \includegraphics[scale=0.7]{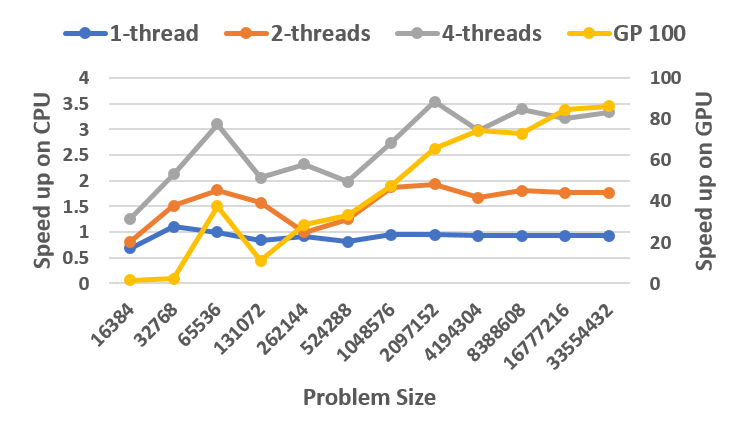}
    \label{fig:blackscholes_nocompile}
  }
  \subfloat[][KMeans]{
    \includegraphics[scale=0.7]{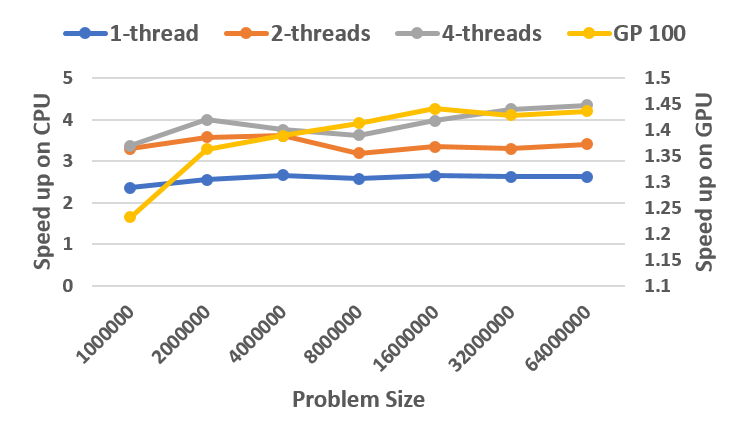}
    \label{fig:kmeans_nocompile}
  }
  \\
  \subfloat[][Jacobian Stencil]{
    \includegraphics[scale=0.7]{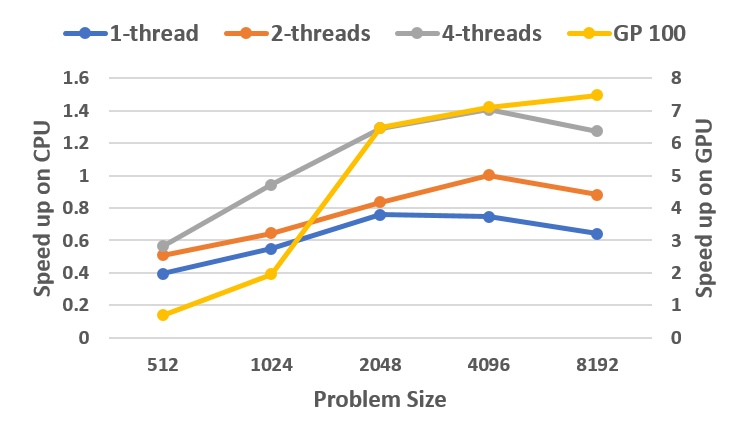}
    \label{fig:jacobi_nocompile}
  }
  %% \subfloat[][Closest Distance on Sphere]{
  %%   \includegraphics[scale=0.7]{figs/cds_nocompile.png}
  %%   \label{fig:cds_nocompile}
  %% }
  %% \\
  \subfloat[][RNN]{
    \includegraphics[scale=0.7]{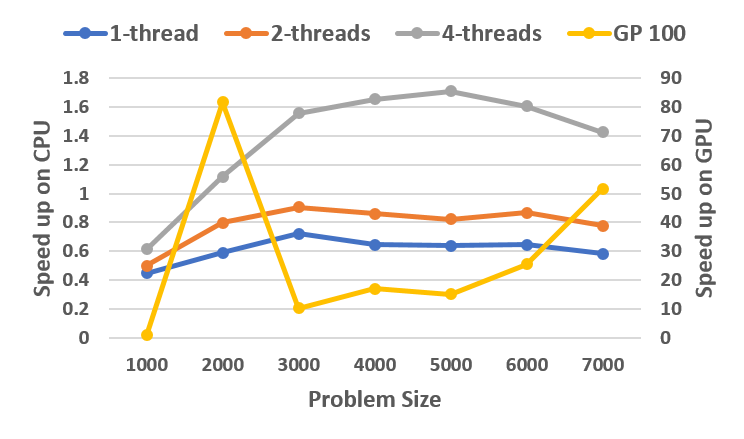}
    \label{fig:rnn_nocompile}
  }
  \caption{Performance comparison against Numpy on multicore CPU and
    NVIDIA GPU without JIT overhead}
  \label{fig:perfbase-nocompile}
\end{figure*}

Another interesting metric to look at is the performance benefit when
the JIT compilation overhead is not accounted for. This would be an
estimate of the benefit from fusion and library usage when compared to
Numpy. For each of the benchmark, Figure~\ref{fig:perfbase-nocompile}
shows the speed up over Numpy when the JIT compilation overhead is not
included in the overall execution time for Grumpy. In general, one
would expect that for small problem sizes the JIT compilation overhead
is large, and this is shown to be the case. The JIT overhead seems to
affect RNN (Figure~\ref{fig:rnn_nocompile}) the most due to high
startup cost, and many small kernels compiled.

\section{Related Work}
\label{sec:related}

Previous works have targeted a similar
approach. Bohrium~\cite{bohrium} offers similar support for
accelerating Numpy frameworks on CPU and GPUs. It effectively supports
lazy evaluation by use of a \emph{GPU Vector Engine} that aggregates
operations into a single CUDA kernel. This effectively implements lazy
evaluation, but does not seem to have a DAG IR that can allow further
optimizations like CSE at the level of array expressions. While such
operations are not currently implemented in Grumpy, having an explicit
DAG IR opens the possibility to do more complex optimization on array
expressions.

ArrayFire~\cite{arrayfire} provides a similar interface to array
computing as Numpy and targets multiple architectures as well. Since
it doesn't support Numpy directly, any Numpy application would have to
be manually ported to this framework. CuPy~\cite{cupy} is listed as a
drop-in replacement of Numpy, but fusion is only possible through user
defined functions through special API. Theano~\cite{theano} with
PyCUDA provides a similar interface. Unlike these frameworks, Grumpy
and Bohrium are true drop-in replacement for Numpy that automatically
generate fused kernels for element-wise operations.

There have also been previous work that support a similar execution
model specifically for Deep Learning applications.  XLA~\cite{xla} is
a compiler framework within Tensorflow that intercepts operations in
Tensorflow to build a graph representation of the operation sequence
and JIT compiles them. These meta operations are then inserted back
into the TensorFlow graph for execution. While it addresses many of
the similar concerns, the focus of this work has mainly been
Deep-Learning applications. NNVM~\cite{nnvm} is a compiler framework
similar to XLA for MxNET. PyTorch~\cite{pytorch} has recently added a
JIT compilation support as well. Being specific to deep-learning the
space of computations supported by these are not as general as
application written in Numpy. %% For example, unlike Grumpy there is no
%% direct support for handling map-scan type of computation in these JIT
%% compilation frameworks.

\section{Future Work}
\label{sec:future}

In this paper we have discussed Grumpy, a drop-in replacement for
Numpy that can automatically build a representation of operations
specified in a Numpy application. This representation is then used to
fuse operations and execute them either on multicore CPUs or GPUs. For
operations where a hand-tuned implementation is available these are
used automatically. While the performance evaluation shown in
Section~\ref{sec:results} show considerable promise, there are several
bottlenecks that need to be addressed.

The main bottleneck is the cost of JIT compilation. Strategies that
hide this cost need to be incorporated into Grumpy. One way to do this
is through pipelining. The entire computation is not executed as a
single fused function in Grumpy, but rather a sequence of
functions. It is possible to overlap the compilation with the
execution of these functions.

The generated code could also be made more efficient. The Grumpy
generated code doesn't employ machine specific optimization like use of
vector instructions on CPUs or shared memory on GPUs. On CPUs, Grumpy
doesn't make use of efficient BLAS libraries either.%%  These could
%% provide more benefit in cases where the JIT compilation overhead is
%% well hidden.

%% Finally, Grumpy uses libraries to execute operations like
%% matrix-matrix multiply on GPUs. Instead it would be useful for the
%% compiler to contain a skeleton of these computations internally. These
%% skeletons could be used to fuse these operations with other
%% operations. For example, fusing pointwise functions that use the
%% result of a matrix-matrix multiply with the operation itself would
%% further reduce the memory usage and memory bandwidth requirements of
%% the kernel.

\bibliographystyle{ACM-Reference-Format}
\bibliography{refs}

\end{document}